\newcommand{\al}{\mbox{$^{26}$\hspace{-0.2em}Al}}

\newcommand{\pcmq}{\mbox{cm$^{-2}$}}

\newcommand{\psec}{\mbox{s$^{-1}$}}
\newcommand{\psr}{\mbox{sr$^{-1}$}}

\newcommand{\fster}{\mbox{ph \pcmq \psec \psr}}

\newcommand{\mrem}{\mbox{\tt spi\_obs\_mrem}}
\def\MeV{\mbox{Me\hspace{-0.1em}V}}

\def\deg{\ensuremath{^\circ}}
\newcommand{\bpar}{\mbox{$\beta_{1},..,\beta_{m}$}}
\newcommand{\acc}{\mbox{$\lambda^k$}}
\newcommand{\um}{\mbox{$\mu$m}}
\def\la{\mathrel{\mathchoice {\vcenter{\offinterlineskip\halign{\hfil
$\displaystyle##$\hfil\cr<\cr\sim\cr}}}
{\vcenter{\offinterlineskip\halign{\hfil$\textstyle##$\hfil\cr
<\cr\sim\cr}}}
{\vcenter{\offinterlineskip\halign{\hfil$\scriptstyle##$\hfil\cr
<\cr\sim\cr}}}
{\vcenter{\offinterlineskip\halign{\hfil$\scriptscriptstyle##$\hfil\cr
<\cr\sim\cr}}}}}
\def\ga{\mathrel{\mathchoice {\vcenter{\offinterlineskip\halign{\hfil
$\displaystyle##$\hfil\cr>\cr\sim\cr}}}
{\vcenter{\offinterlineskip\halign{\hfil$\textstyle##$\hfil\cr
>\cr\sim\cr}}}
{\vcenter{\offinterlineskip\halign{\hfil$\scriptstyle##$\hfil\cr
>\cr\sim\cr}}}
{\vcenter{\offinterlineskip\halign{\hfil$\scriptscriptstyle##$\hfil\cr
>\cr\sim\cr}}}}}

          [2001/04/25 1.1 (PWD)]
\documentclass[a4paper,twocolumn]{esapub2005} 
\usepackage{times}
\usepackage{natbib}
\usepackage{graphicx}

\title{Imaging the Gamma-Ray sky with SPI aboard INTEGRAL}
\author{J\"urgen Kn\"odlseder}
\author{G. Weidenspointner}
\author{P. Jean}
\affil{Centre d'\'Etude Spatiale des Rayonnements, CNRS/UPS, B.P.~4346, 
       31028 Toulouse Cedex 4, France}
\author{R. Diehl}
\author{A. Strong}
\affil{Max-Planck-Institut f\"ur Extraterrestrische Physik, Postfach 1603, 
       85740 Garching, Germany}
\author{H. Halloin}
\affil{APC, 75231 Paris Cedex 5, France}
\author{B. Cordier}
\author{S. Schanne}
\affil{CEA Saclay, DSM/DAPNIA/Service d'Astrophysique, 91191 Gif-sur-Yvette, 
       France}
\author{C. Winkler}
\affil{ESA/ESTEC, Science Operations and Data Systems Division 
       (SCI-SD), 2201 AZ Noordwijk, The Netherlands}

\begin{document}


\maketitle

\begin{abstract}
The spectrometer SPI on INTEGRAL allows for the first time simultaneous 
imaging of diffuse and point-like emission in the hard X-ray and soft
gamma-ray regime. 
To fully exploit the capabilities of the instrument, we implemented the 
MREM image deconvolution algorithm, initially developed for COMPTEL data 
analysis, to SPI data analysis. 
We present the performances of the algorithm by means of simulations 
and apply it to data accumulated during the first 2 mission years of 
INTEGRAL.
Skymaps are presented for the 1809 keV gamma-ray line, attributed to 
the radioactive decay of \al, and for continuum energy bands, 
covering the range 20 keV -- 3 \MeV.
The 1809 keV map indicates that emission is clearly detected by SPI 
from the inner Galactic radian and from the Cygnus region.
The continuum maps reveal the transition between a point-source 
dominated hard X-ray sky to a diffuse emission dominated soft 
gamma-ray sky.
From the skymaps, we extract a Galactic ridge emission spectrum that 
matches well SPI results obtained by model fitting 
\cite{bouchet05,strong05}.
By comparing our spectrum with the cumulative flux measured by IBIS 
from point sources, we find indications for the existence of an 
unresolved or diffuse emission component above $\sim$100 keV.
\end{abstract}

\section{Introduction}

The image deconvolution of gamma-ray data is confronted with several 
problems that are specific to the domain.
First, the data are dominated by a strong instrumental background, 
generated by the interaction of cosmic-rays with the telescope 
detectors and the surrounding satellite materials.
For SPI, typical signal-to-background ratios are of the order of
$\sim$1\%.
Hence, the deconvolution seeks to explain only a tiny fraction of the 
registered events, while the bulk of the data have to be explained 
by a model of the instrumental background characteristics.
Second, the information measured by the telescope for each registered 
photon is only indirectly related to its arrival direction.
For SPI, the general data space is 3-dimensional, spanned by the 
pointing number, the detector identifier, and the measured energy.
Analysing the data in a single energy band reduces the date space to two 
dimensions, which differ however considerably in nature from the 
image space which is spanned by longitude and latitude.
Third, measured count rates are generally very low, leading to non 
Gaussian distributions for the statistical measurement uncertainties.
The consideration of the Poissonian nature of the data is in many cases 
mandatory to derive meaningful error estimates. 

All these particularities explain why imaging the gamma-ray sky with SPI
aboard INTEGRAL is difficult, and why classical image reconstruction 
methods rapidly show their limitations in this domain.
Statistical noise easily propagates into the reconstructed sky 
intensity distributions, leading to artefacts that can severely
mislead the observer.
One may seek to suppress this statistical noise using regularisation 
methods, premature stopping of the iterations, and/or image smoothing
\cite{strong03,knoedl05,allain06}.
These tricks go however in general at the expense of flux suppression 
and a loss in the angular resolution.

We discuss in this paper the application of an alternative method for 
image deconvolution of INTEGRAL/SPI data which is based on a 
multiresolution analysis of the data using wavelets.
The method is an implementation of the Multiresolution Expectation 
Maximisation (MREM) algorithm that has been initially developed to 
perform image deconvolution of gamma-ray data that where acquired 
by the COMPTEL telescope aboard the Compton Gamma-Ray Observatory 
(CGRO) \cite{knoedl99a}.
The algorithm is implemented as the program \mrem\
(version 3.4.0)
and is publicly available at the web site
{\tt http://www.cesr.fr/$\sim$jurgen/isdc}.
We will demonstrate that MREM is equally well suited for imaging of 
point source and diffuse emission, that the algorithm is able to cope 
with large exposure gradients over the imaged field, and that it allows 
at the same time reliable estimates of gamma-ray intensities from the 
resulting images.
MREM is a convergent algorithm, and it will be shown that its only 
sensitive parameter is the wavelet thresholding level which has a 
well defined scaling, allowing for an objective a priori choice of its 
value.

\section{Deconvolution algorithms}

\subsection{Richardson-Lucy algorithm}

The starting point of MREM is the Richardson-Lucy (RL) algorithm
\cite{richardson72, lucy74}
which is widely used for the deconvolution of astronomical images, 
and which, in particular, has been also successfully employed for the 
analysis of gamma-ray data of CGRO 
\cite{knoedl99a, milne00}.
Starting from an initial estimate $f_j^0$ for the image
which we usually take as a grey image of negligible 
intensity\footnote{
This choice corresponds to having no a priori assumption about the 
possible spatial distribution of the emission.},
RL iteratively improves this estimate using the relation
\begin{equation}
 f_j^{k+1} = f_j^k + \delta_j^k
\label{eq:rl}
\end{equation}
with
\begin{equation}
 \delta_j^k = f_j^k \times
	      \sum_{i=1}^N \left( \frac{n_{i}}{e_{i}^k} - 1 \right)  R_{ij}
 \label{eq:delta}
\end{equation}
being the image increment of iteration $k$ and 
\begin{equation}
 e_{i}^k = \sum_{j=1}^M R_{ij} f_j^k + b_{i}
 \label{eq:expect}
\end{equation}
being the predicted number of counts in data space bin $i$ after 
iteration $k$, where
\begin{itemize}
\item $i$ is the index of the data space, running from $1$ to $N$,
\item $j$ is the index of the image space, running from $1$ to $M$,
\item $k$ counts the number of iterations, starting with $0$ for the 
      initial image estimate,
\item $f_j^k$ if the intensity in image pixel $j$ after iteration $k$
      (units: \fster),
\item $R_{ij}$ is the instrumental response matrix linking the data space 
      to the image space (units: counts ph$^{-1}$ cm$^2$ s sr),
\item $n_{i}$ is the number of counts measured in data space bin $i$
      (units: counts),
\item $b_{i}$ is the predicted number of instrumental background counts 
      for data space bin $i$ (units: counts).
\end{itemize}
Convergence of this algorithm may eventually be very slow, in 
particular if the instrumental background, $b_{i}$, is the dominating 
term in Eq.~(\ref{eq:expect}).
Further, the model of the instrumental background may be 
presented in a parametrised form, e.g.
$b_{i}(\bpar)$,
where the background model parameters \bpar\ are to be determined 
along with the image deconvolution process.

Convergence acceleration and simultaneous background model parameter 
fitting is achieved by replacing Eq.~(\ref{eq:rl}) by
\begin{equation}
 f_j^{k+1} = f_j^k + \lambda^k \delta_j^k .
\label{eq:linb}
\end{equation}
and by optimising the acceleration factor $\lambda^k$ together with 
the background model parameters \bpar\ by maximising the log-likelihood 
function
\begin{equation}
 L(\acc,\bpar) = \sum_{i=1}^N n_i \ln e_{i}^{k+1} - e_{i}^{k+1} - 
     \ln n_i !
\label{eq:lm}
\end{equation}
with
\begin{equation}
 e_{i}^{k+1} = \sum_{j=1}^M R_{ij} (f_j^k + \acc \delta_j^k) + 
 b_{i}(\bpar) .
 \label{eq:model}
\end{equation}
To conserve positivity of the image pixels during the acceleration 
process (a property which is inherent to the standard RL algorithm) we 
require $\lambda^k \delta_j^k > -f_j^k$ for all image pixels $j$.

\subsection{MREM algorithm}

The RL algorithm as implemeted by 
Eqs.~\ref{eq:linb}, \ref{eq:delta}, \ref{eq:lm}, and \ref{eq:model}
does not explicitely take into account the correlation of intensities 
in neighbouring image pixels.
In the case of weak signals, individual pixels are therefore poorly 
constrained by the data and statistical fluctuations easily propagate 
in the reconstruction and quickly amplify.
One approach to limit this noise amplification is the application of a 
smoothing kernel to the image increments $\delta_j^k$, i.e. to replace
\begin{equation}
 \delta_j^k \to \sum_{r=1}^M \Pi_{jr} \delta_r^k .
 \label{eq:smoothing}
\end{equation}
As an example, the 511 keV allsky map presented in \cite{knoedl05} 
has been obtained by the RL algorithm using a 5\deg$\times$5\deg\ boxcar 
average.
Together with premature stopping of the iterations, this resulted in a 
basically noise free image of the Galactic 511 keV intensity 
distribution.

The drawback of this approach is 
(1) the effective limited angular resolution imposed by the smoothing 
procedure and 
(2) the arbitrary stopping of the iterations.
In the MREM approach, we circumvent these problems by 
replacing Eq.~\ref{eq:smoothing} by the wavelet filtering step
\begin{equation}
 \delta_j^k \to \sum_{s=1}^{M'} W^{-1}_{sj} \eta_s^k c_s^k
 \label{eq:wavelet}
\end{equation}
where
\begin{equation}
 c_s^k = \sum_{j=1}^M W_{sj} \delta_j^k
 \label{eq:wcoeff}
\end{equation}
are the wavelet coefficients of the image increment,
$W_{sr}$ is a $M' \times M$ matrix that present the wavelet 
transform (and $W^{-1}_{sj}$ the inverse transform), and 
\begin{equation}
 \eta_s^k = 
 \left\{
 \begin{array}{r@{\quad:\quad}l}
     0 & c_s^k < t_s^k \\
     1 & \mbox{else}
 \end{array}
 \right.
\end{equation}
is a threshold vector for the wavelet coefficients $c_s^k$.
The threshold vector represents the essence of the filtering step in 
that it zeros all those wavelet coefficients that fall below a given 
threshold $t_s^k$.
The idea behind this kind of filtering is that significant image 
structures are represented by large wavelet coefficients while 
insignificant structures are represented by small coefficients.
The wavelet transform operates globally on the image and therefore 
introduces pixel-to-pixel correlations on all possible scales, 
and the careful selection of $t_s^k$ allows us to keep only 
structures for which significant evidence is present in the data.
Once all significant structures have been extracted, the algorithm 
stops.

\begin{figure*}[!t]
\centering
\includegraphics[width=0.86\linewidth]{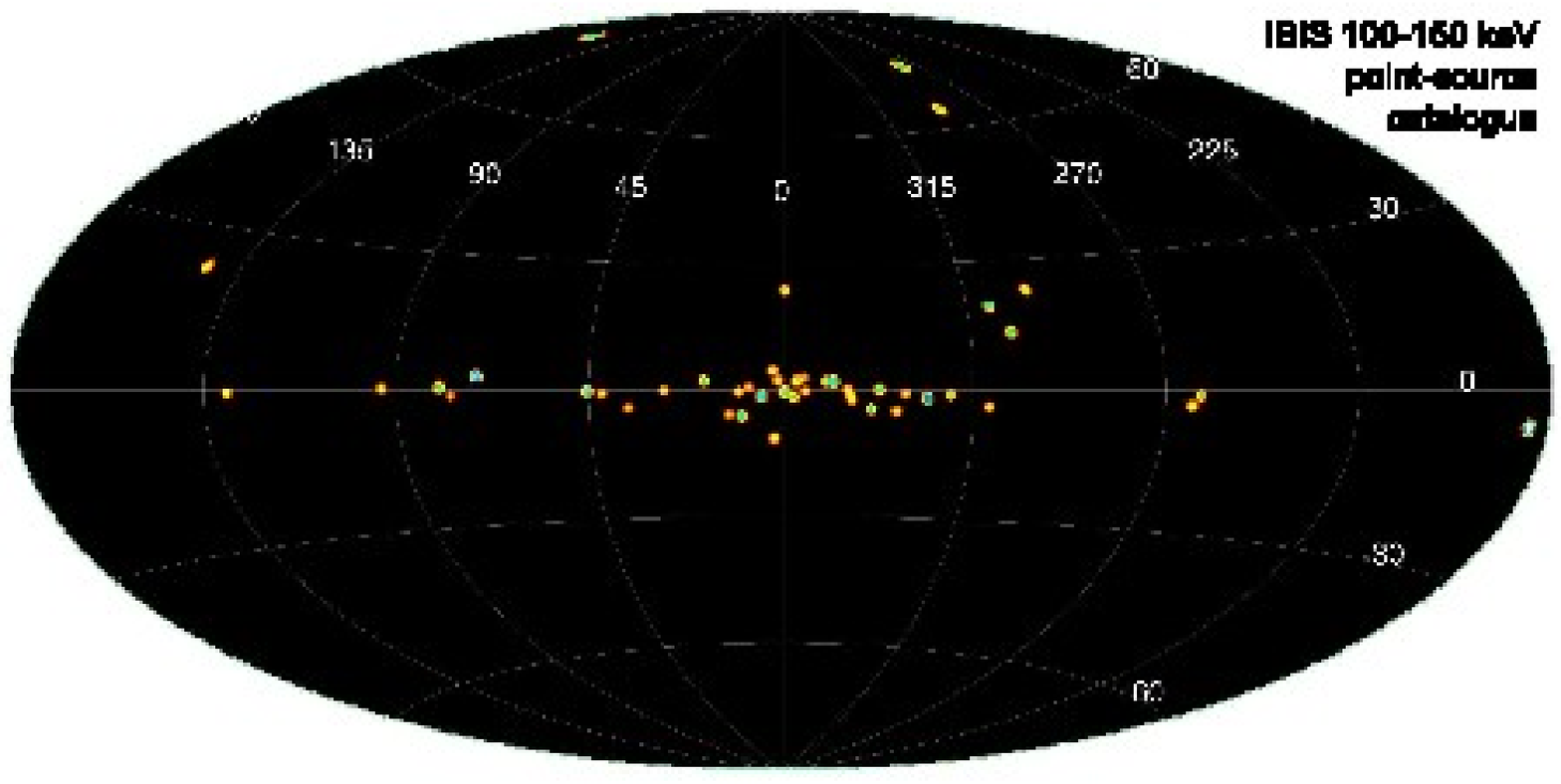}
\includegraphics[width=0.86\linewidth]{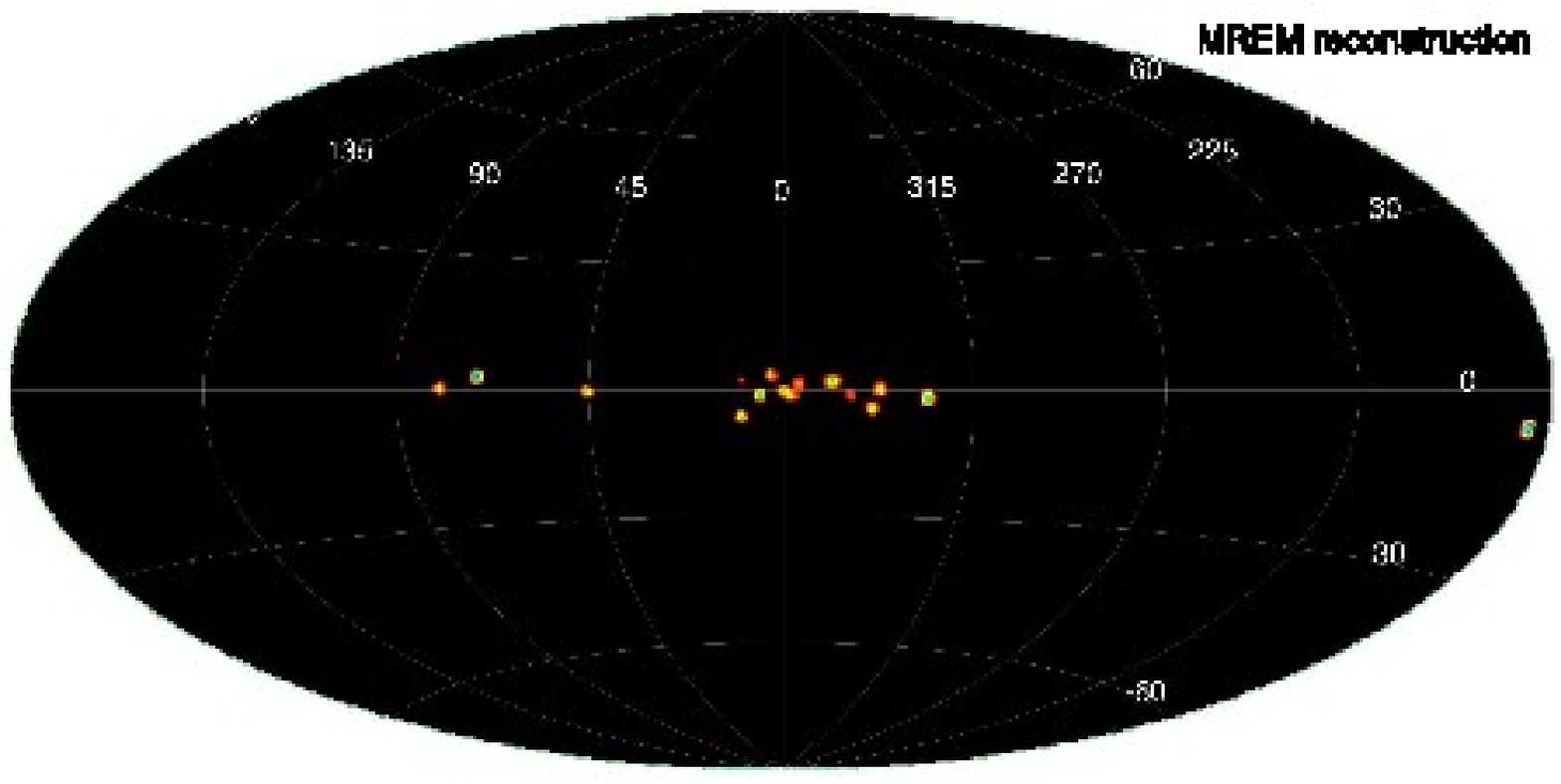}
\caption{
  Model (top) and MREM reconstruction (bottom) of a mock dataset of 
  point sources.
  The simulated sky model is a compilation of IBIS point-sources detected 
  in the 100--150 keV band \cite{bazzano06}.
  For the purpose of better visibility, the emission for each source 
  in the model has been spread over a diameter of 2 degrees around 
  the source position.
  Image reconstruction has been performed using the SPI response in 
  the 100--150 keV energy band.
  \label{fig:sim-point}}
\end{figure*}

We estimate $t_s^k$ for each iteration using Monte-Carlo simulations.
For this purpose we create simulated image increments by replacing in 
Eq.~\ref{eq:delta} the measured number of counts, $n_i$, by the mock 
data $\tilde{n_i}$ which have been drawn from the current estimates
$e_i^k$ using a Poisson random number generator.
Since $\tilde{n_i}$ are a possible realisation of $e_i^k$, the 
resulting image increment reflects now the impact of statistical 
noise fluctuations only.
Transforming this fake image increment in the wavelet domain using 
Eq.~\ref{eq:wcoeff} gives then possible values for all wavelet 
coefficients in the presence of pure statistical noise.
Repeating this process a sufficient number of times allows then to 
estimate the root-mean-square, $\sigma_s^k$, for each coefficient.
We use then these estimates and set the thresholds to
\begin{equation}
 t_s^k = \alpha \times \sigma_s^k
\end{equation}
where $\alpha$ may be interpreted as the number of standard deviations 
of noise that are allowed during the reconstruction.
We found satisfactory results for $\alpha \approx$3.5. 
100 Monte Carlo simulations per iteration were found sufficient for 
reliably estimating the noise levels $\sigma_s^k$.

\section{Simulations}
\label{sec:sim}

\begin{figure*}[!t]
\centering
\includegraphics[width=0.86\linewidth]{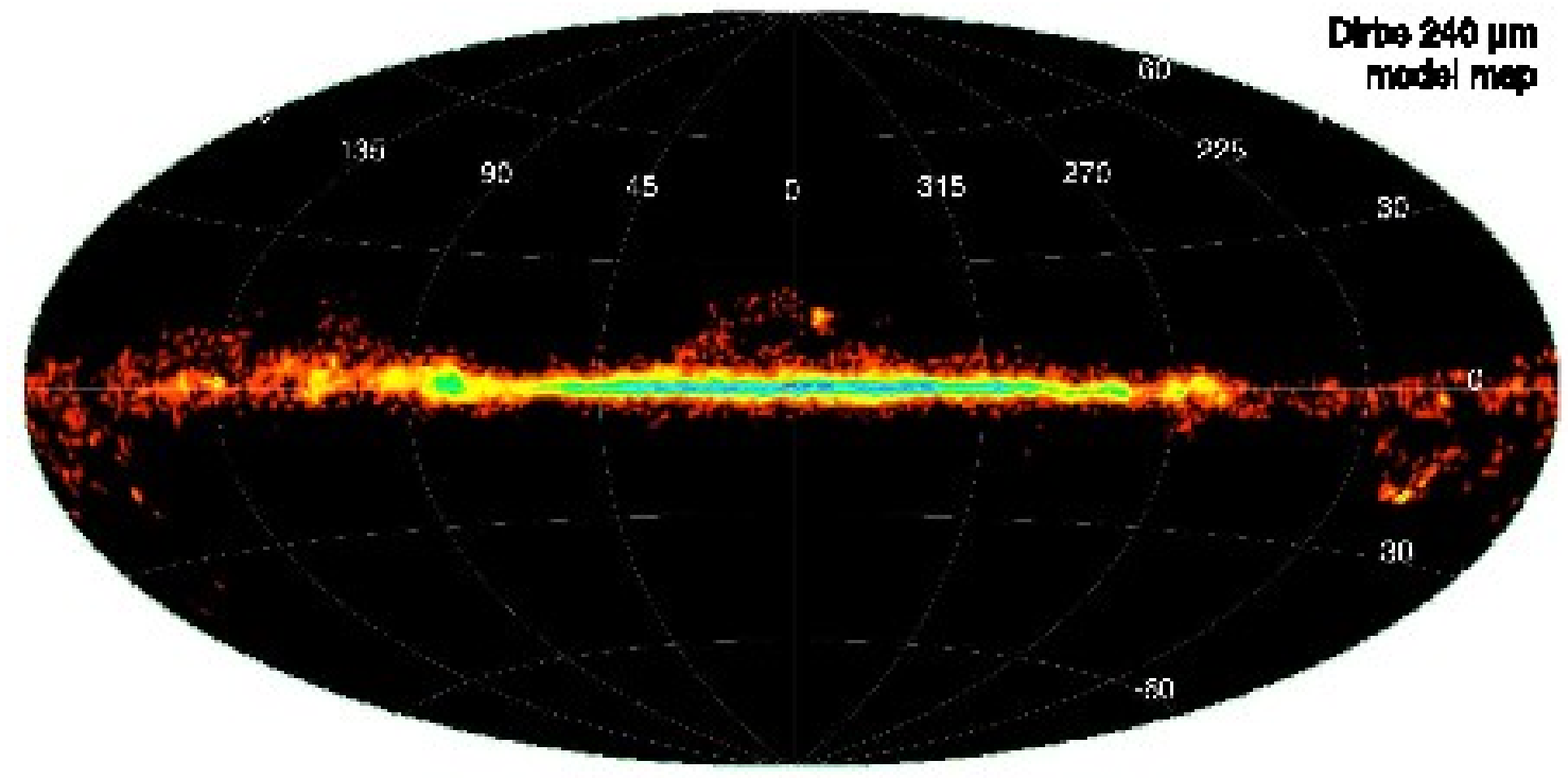}
\includegraphics[width=0.86\linewidth]{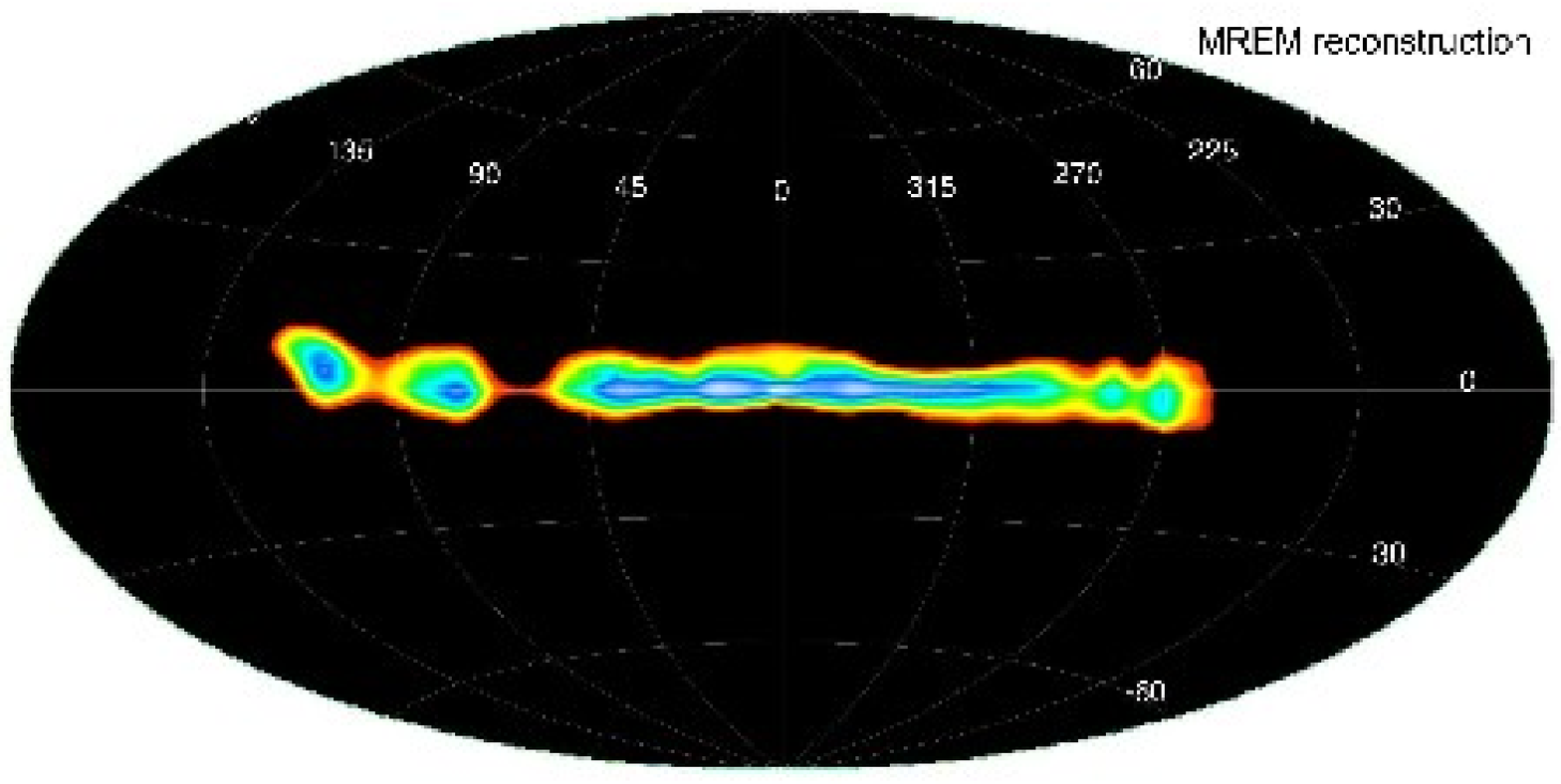}
\caption{
  Model (top) and MREM reconstruction (bottom) of a mock dataset of 
  diffuse emission.
  The simulated sky model is the Dirbe 240 \um\ cold dust emission 
  map, scaled to an intensity that corresponds to about 5 times the 
  intensity of the Galactic 1809 keV line emission.
  Image reconstruction has been performed using the SPI response in 
  a 7 keV wide energy band centred on 1808.5 keV.
  \label{fig:sim-diffuse}}
\end{figure*}

To illustrate the performances of MREM we present two simulated 
reconstructions in Figs.~\ref{fig:sim-point} and \ref{fig:sim-diffuse}.
Both simulations were based on the SPI pointing sequence of the first 
2 years of observations, spanning the orbital revolutions 19--269.
In both cases, we adopt a realistic model for the instrumental 
background that was based on the measured data.
To simulate the gamma-ray sky, we convolve model intensity 
distributions with the instrumental response, and add the resulting 
data space model to that of the instrumental background.
Mock data are then generated from these models using our tool 
{\tt spi\_obs\_sim}, which employs a Poisson random number generator 
to create a realisation of measured events that is statistically 
compatible with the model.

Our first example illustrates the MREM reconstruction of point source 
emission (Fig.~\ref{fig:sim-point}).
As sky model we used the IBIS catalogue of 49 sources found in the 
100--150~keV energy band \cite{bazzano06}.
The mock dataset has been created and the MREM reconstruction has been 
performed for the same energy band, but now for the SPI instrument 
which is less sensitive with respect to IBIS.
Consequently, from the 49 simulated sources MREM reaveals only the 15 
most brightest ones in the reconstruction.
In addition, the modest angular resolution of SPI of $\sim$3\deg\ 
leads to some source confusion towards the densely populated Galactic 
centre region.
Besides the 15 point sources, no other emission features are present, 
and in particular, the map is free from image noise.

Our second example illustrates the MREM reconstruction of diffuse emission
(Fig.~\ref{fig:sim-diffuse}).
As sky model we've chosen the Dirbe 240 \um\ cold dust emission map 
which has been shown to provide a reasonable representation of the 
spatial distribution of 1809 keV gamma-ray line emission attributed 
to the radioactive decay of \al\ \cite{knoedl99b}.
In order to highlight details of the reconstruction process, we 
scaled the intensity of the skymap to approximately 5 times of that expected 
from 1809 keV line emission.
The simulation and reconstruction was performed for a 7 keV wide energy 
band centred on 1808.5 keV.
MREM clearly picks up the basic features of the input map:
bright and narrow Galactic ridge emission plus several localised emission 
features in the Cygnus, Carina and Vela regions.
Low-intensity emission that is coded in reddish color in the model map 
falls apparently below the SPI detection limit, and consequently is clipped 
by the wavelet filter during the deconvolution.
Also for this case, the reconstruction appears to be free of image noise.

Another interesting feature of the MREM reconstruction is the 
sharpening of the Galactic ridge emission towards the regions of 
high-intensity in the inner Galaxy.
Since the signal in this area is detected at higher statistical 
significance, the data allow to constrain the intensity profile more 
precisely, and the reconstructed profile becomes sharper.
Further away from the Galactic centre, say beyond $\pm$40\deg\ Galactic 
longitude, the weaker signal intensity makes the data less 
constraining, and consequently, the reconstructed profile becomes broader.
This behaviour is well known in statistics as the 
{\em bias-variance trade-off},
which generally occurs when we try to separate signal from noise 
using smoothing or correlation techniques.

\section{1809 keV gamma-ray line emission}
\label{sec:1809keV}

\begin{figure*}[!ht]
\centering
\includegraphics[width=0.86\linewidth]{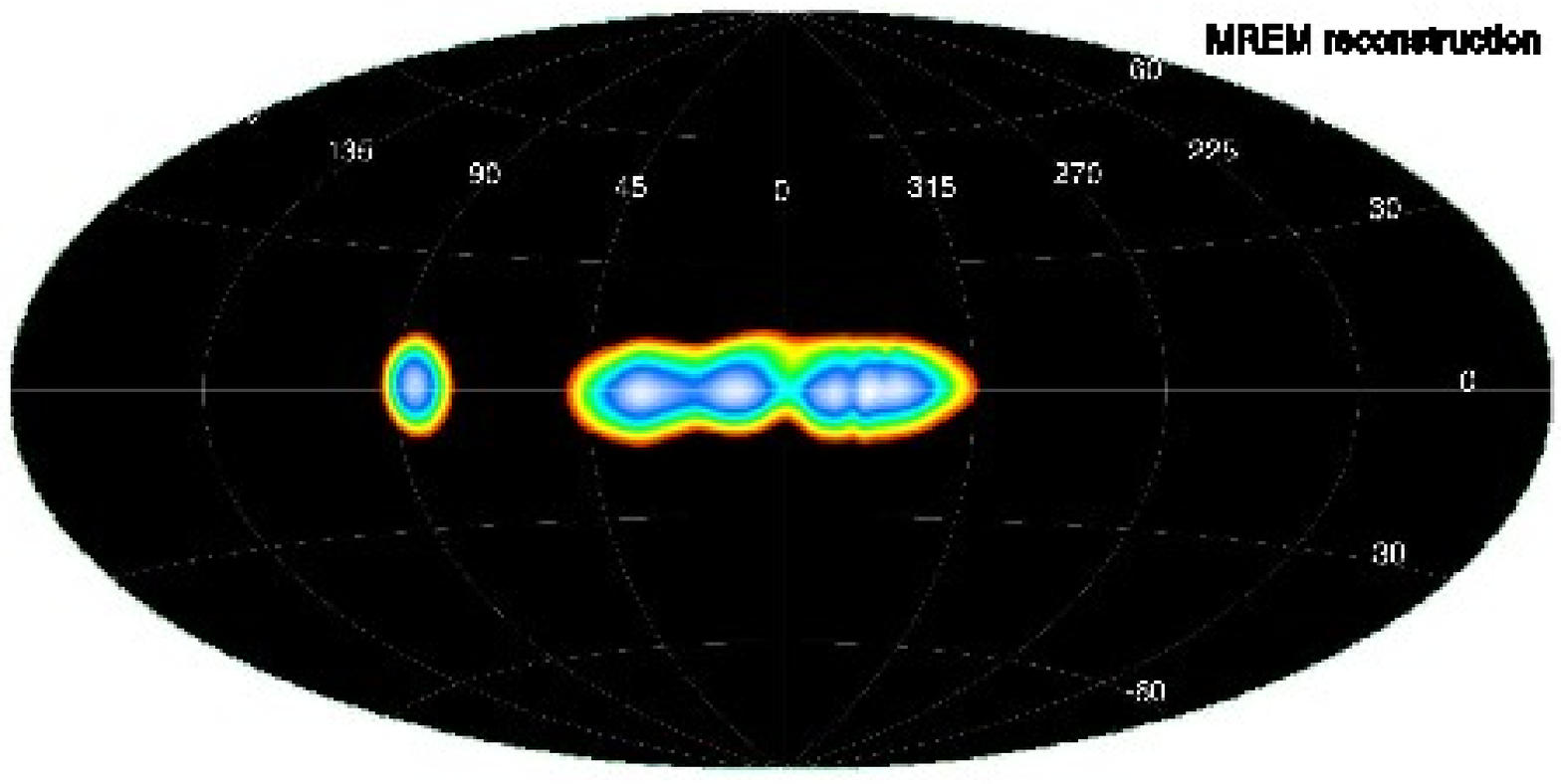}
\includegraphics[width=0.86\linewidth]{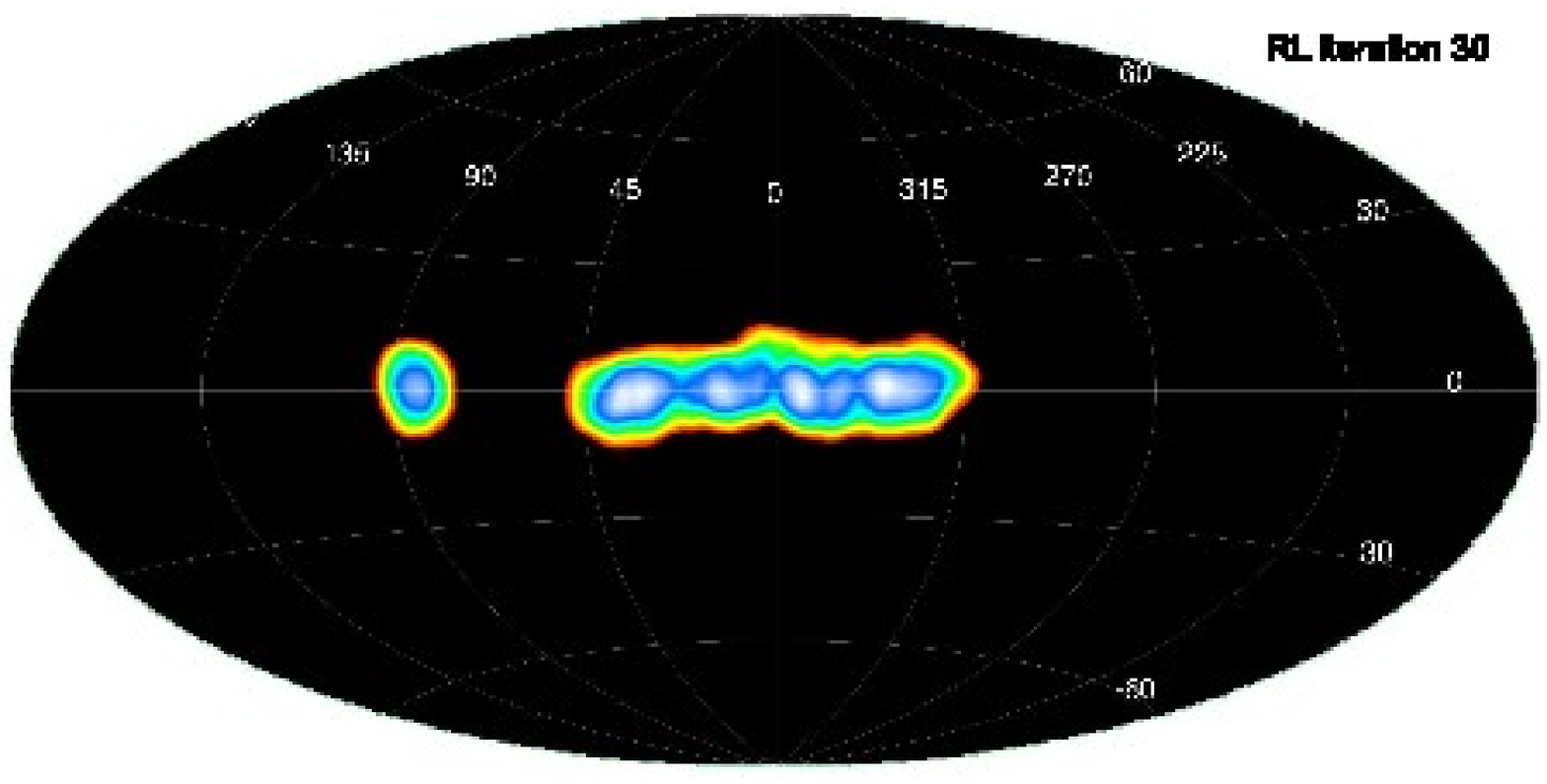}
\includegraphics[width=0.86\linewidth]{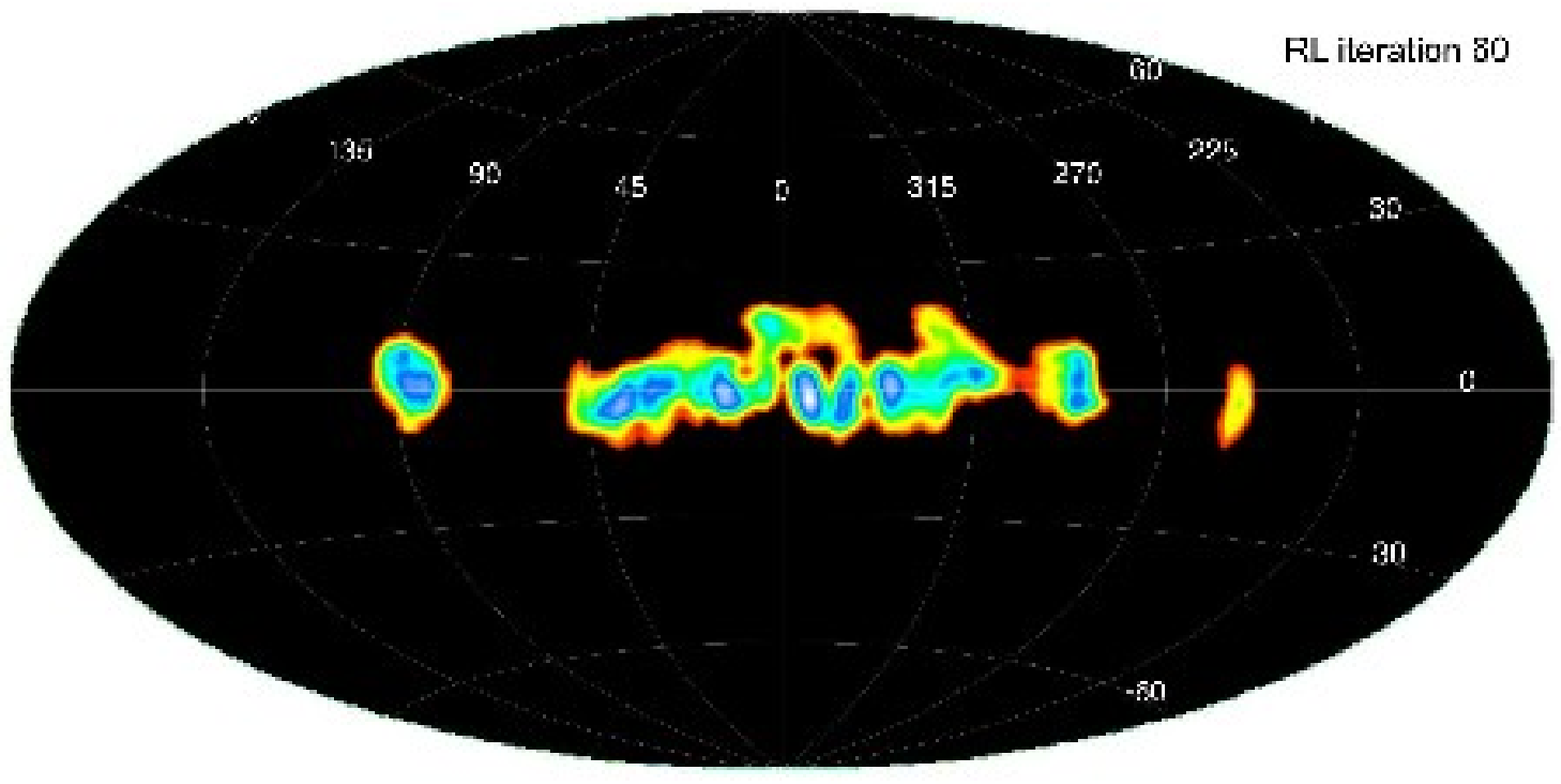}
\caption{
  Comparison of MREM (top panel) and RL (middle and bottom panel) 
  reconstructions of 1809 keV data.
  For RL we show the result after iteration 30 and 60.
  To reduce the image noise, we applied a boxcar average of 
  15\deg$\times$15\deg\ for the RL reconstruction.
  \label{fig:1809keV}}
\end{figure*}

We now apply the MREM algorithm to real data.
The dataset we used is the same as that employed for the simulations:
the first 2 years of INTEGRAL/SPI observations, spanning the orbital 
revolutions 19--269.
As usually in our analyses, we screened the data for solar flares, 
exceptionally high countrates, and other possible problems.

Our first goal is to image the sky in the 1809 keV gamma-ray line 
attributed to the radioactive decay of \al.
Similar to the simulation, we analyse the data in a 7 keV wide energy 
band centred on 1808.5 keV.
To fully exploit the sensitivity of SPI we use both single and double events.
The resulting MREM image is shown in the top panel of Fig.~\ref{fig:1809keV}.
For comparison, we also show the results of the RL reconstruction
after iterations 30 and 60 in the middle and bottom panels of  
Fig.~\ref{fig:1809keV}.
To reduce image noise in the RL reconstruction we applied a boxcar 
average of 15\deg$\times$15\deg\ to smooth the image increment after 
each iteration (cf.~Eq.~\ref{eq:smoothing}).
The images may also be compared to the RL image presented in
\cite{diehl06} which uses an exponential disk as initial image 
estimate $f_j^0$ and for which no boxcar averaging has been performed.

The MREM image indicates that 1809 keV line emission is clearly detected 
by SPI from the inner Galactic radian (say between $\pm$45\deg\ Galactic 
longitude) and from the Cygnus region (situated in the Galactic plane 
at $\sim$80\deg\ longitude).
The inner Galaxy emission shows some local maxima, but those are 
probably of low statistical significance.
Indeed, when we increase the wavelet threshold level from our standard 
value of $\alpha$=3.5 to 4 most of these maxima disappear, 
suggesting that the features were probably linked to statistical noise 
fluctuations.

COMPTEL observations have indicated 1809 keV line emission also from 
the Carina and Vela regions, located along the Galactic plane at 
longitudes $\sim$286\deg\ and $\sim$266\deg, respectively
\cite{knoedl96,diehl95}.
The RL reconstruction shows indeed a hint for emission from Carina 
(visible after iteration 60), but the large amount of structure in the 
image towards the inner Galaxy -- which is due to statistical noise 
fluctuations -- indicates that the significance of the Carina feature 
is weak in our analysed SPI dataset.
So far our imaging analysis shows no indication for 1809 keV line 
emission from the Vela region, but also in the COMPTEL data, the 
statistical significance of this feature was among the weakest
(see also Schanne et al., these proceedings).

\begin{figure*}[!ht]
\centering
\includegraphics[width=0.86\linewidth]{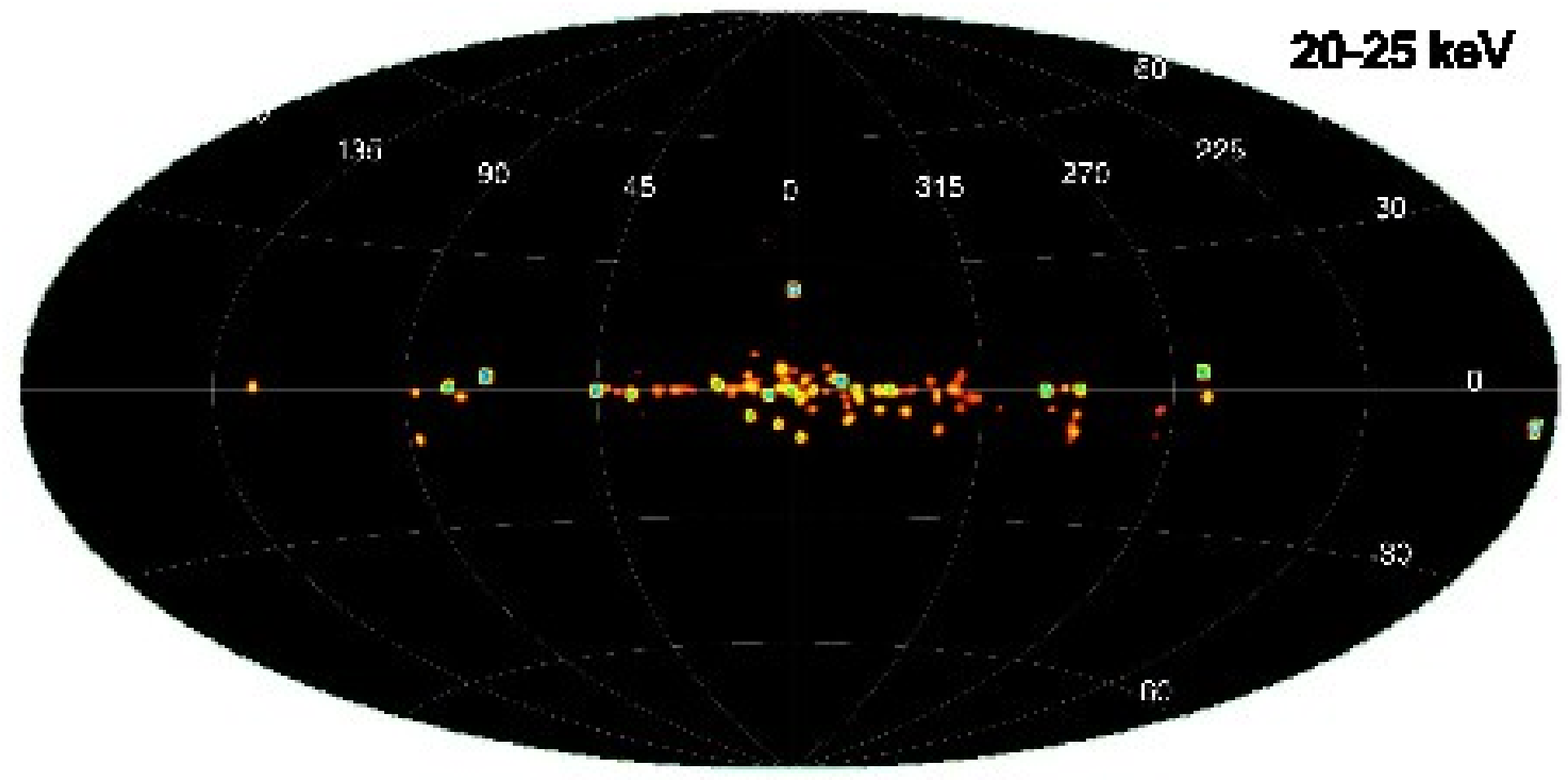}
\includegraphics[width=0.86\linewidth]{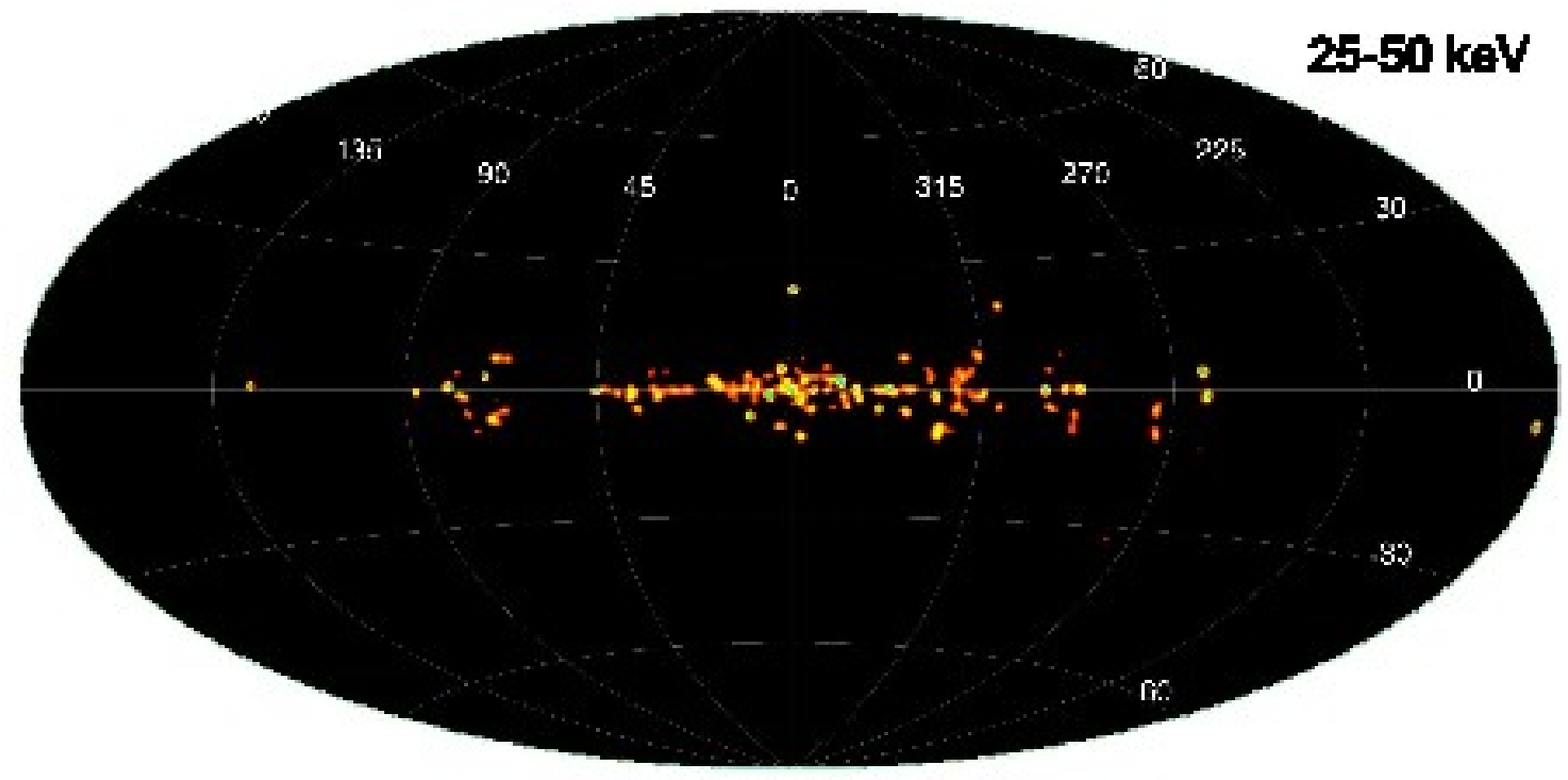}
\includegraphics[width=0.86\linewidth]{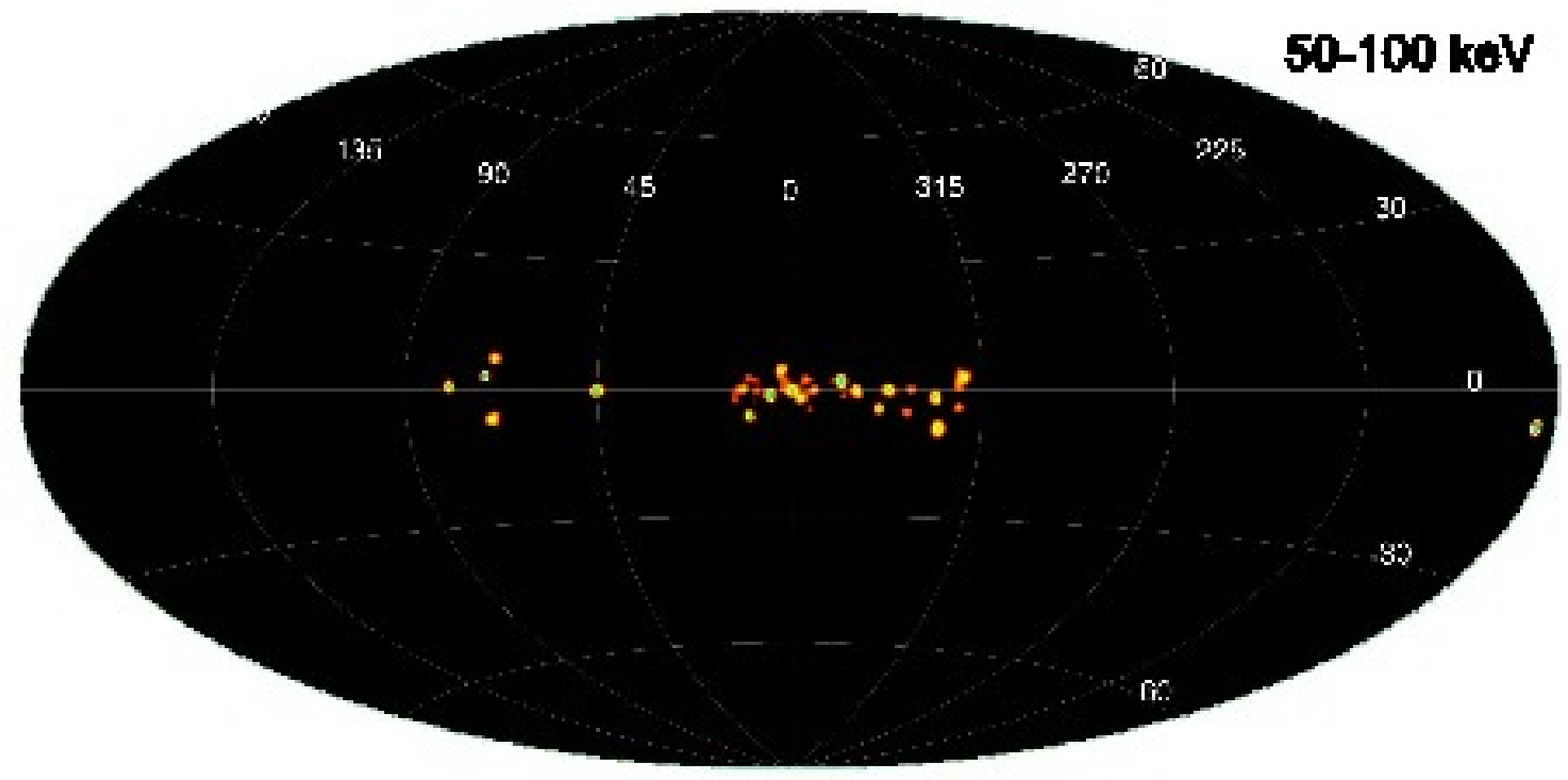}
\caption{
  MREM images of the hard X-ray sky in the continuum energy bands
  20--25, 25--50, and 50--100 keV.
  \label{fig:cont1}}
\end{figure*}

It is interesting to note that the RL image after iteration 30 appears 
similar to the MREM image.
The heavy boxcar smoothing of 15\deg$\times$15\deg\ that we applied 
during the RL deconvolution plays apparently a similar role as the 
wavelet filter in that it suppresses high-frequency image noise during 
the reconstruction.
However, the RL iterations do not stop at this point, and even such 
heavy smoothing does not prevent the growth of image noise in the 
reconstructions with proceeding iterations.
Although we have not formulated an explicit stopping criterion for RL, 
the algorithm did indeed stop after 85 iterations with an image 
similiar to that shown after iteration 60.
In fact, at this point the image increment smoothing did prevent any further 
improvement of the likelihood, so the iterations were aborted.

\begin{figure*}[!ht]
\centering
\includegraphics[width=0.86\linewidth]{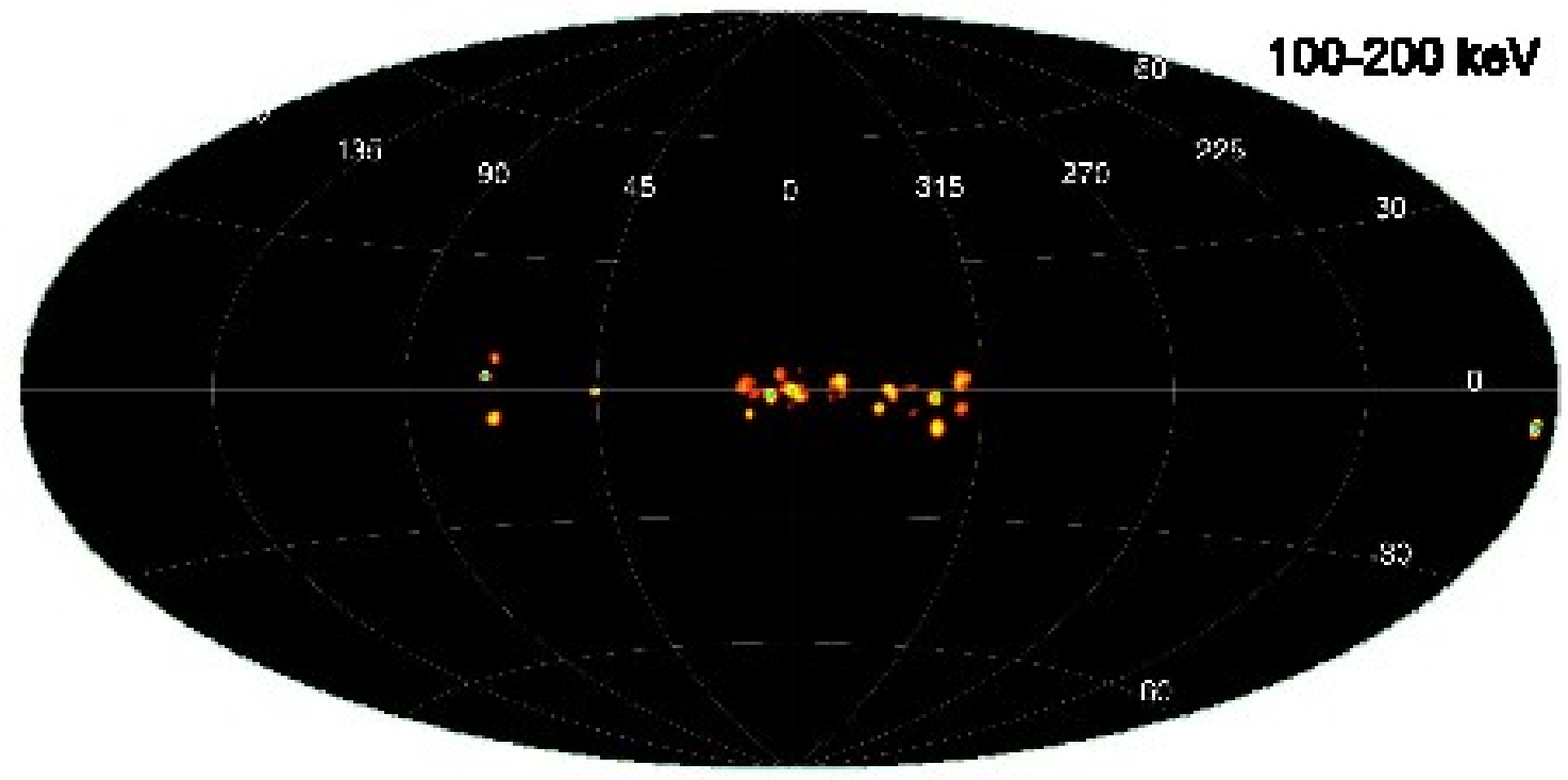}
\includegraphics[width=0.86\linewidth]{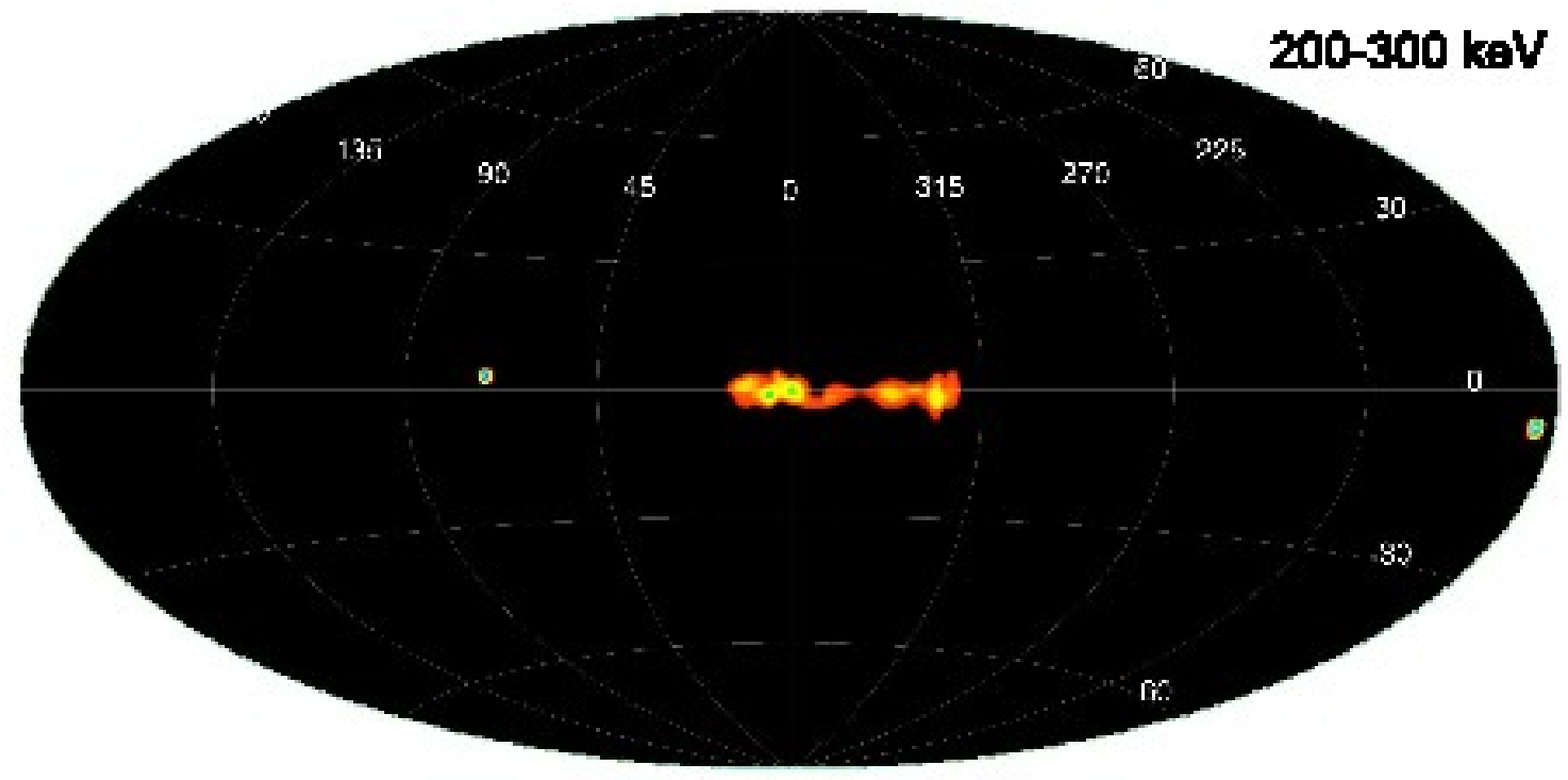}
\includegraphics[width=0.86\linewidth]{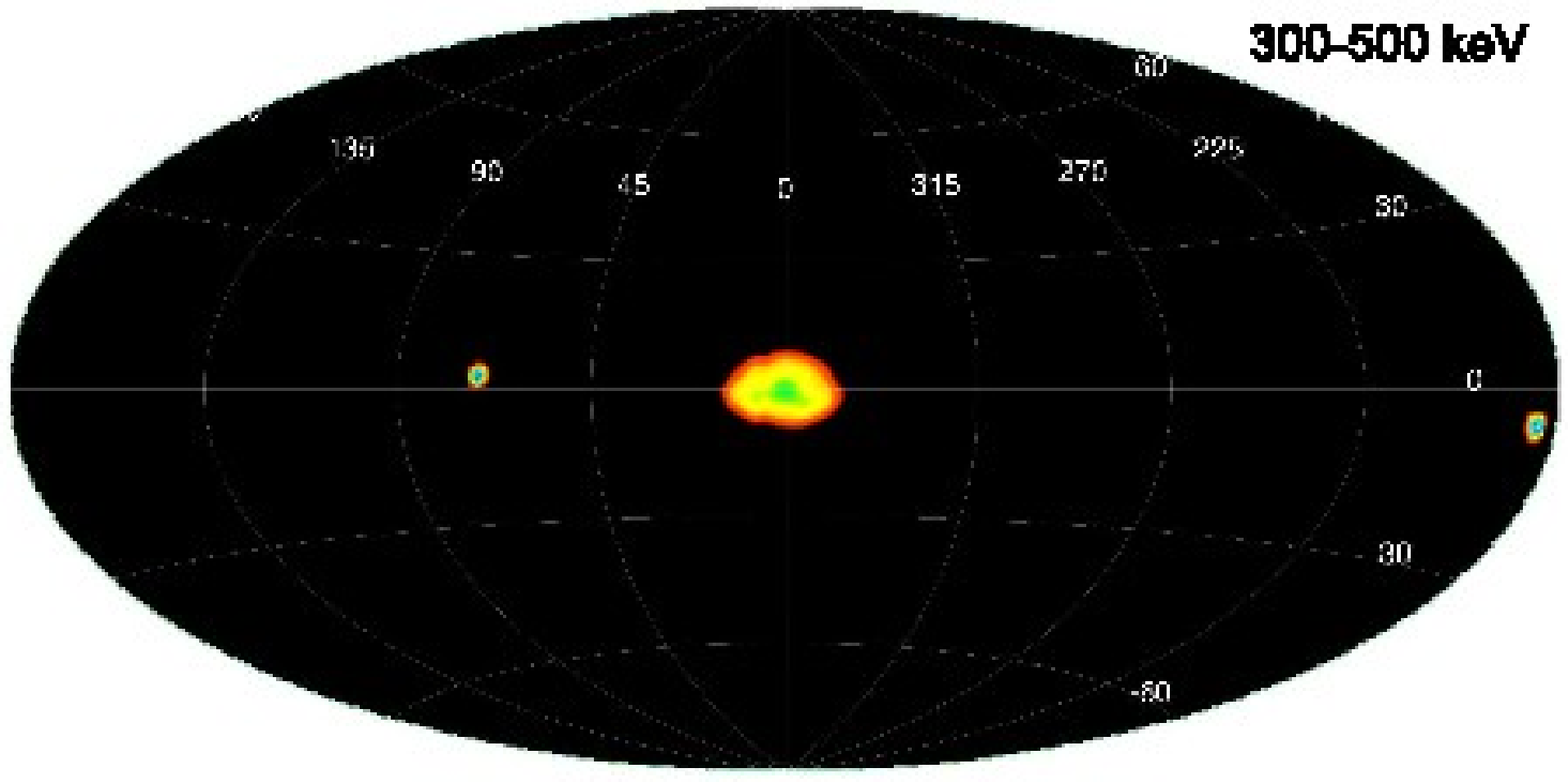}
\caption{
  MREM images of the hard X-ray and soft gamma-ray sky in the continuum
  energy bands 100--200, 200--300, and 300--500 keV.
  \label{fig:cont2}}
\end{figure*}

\section{Galactic continuum emission}
\label{sec:cont}

\begin{figure*}[!ht]
\centering
\includegraphics[width=0.86\linewidth]{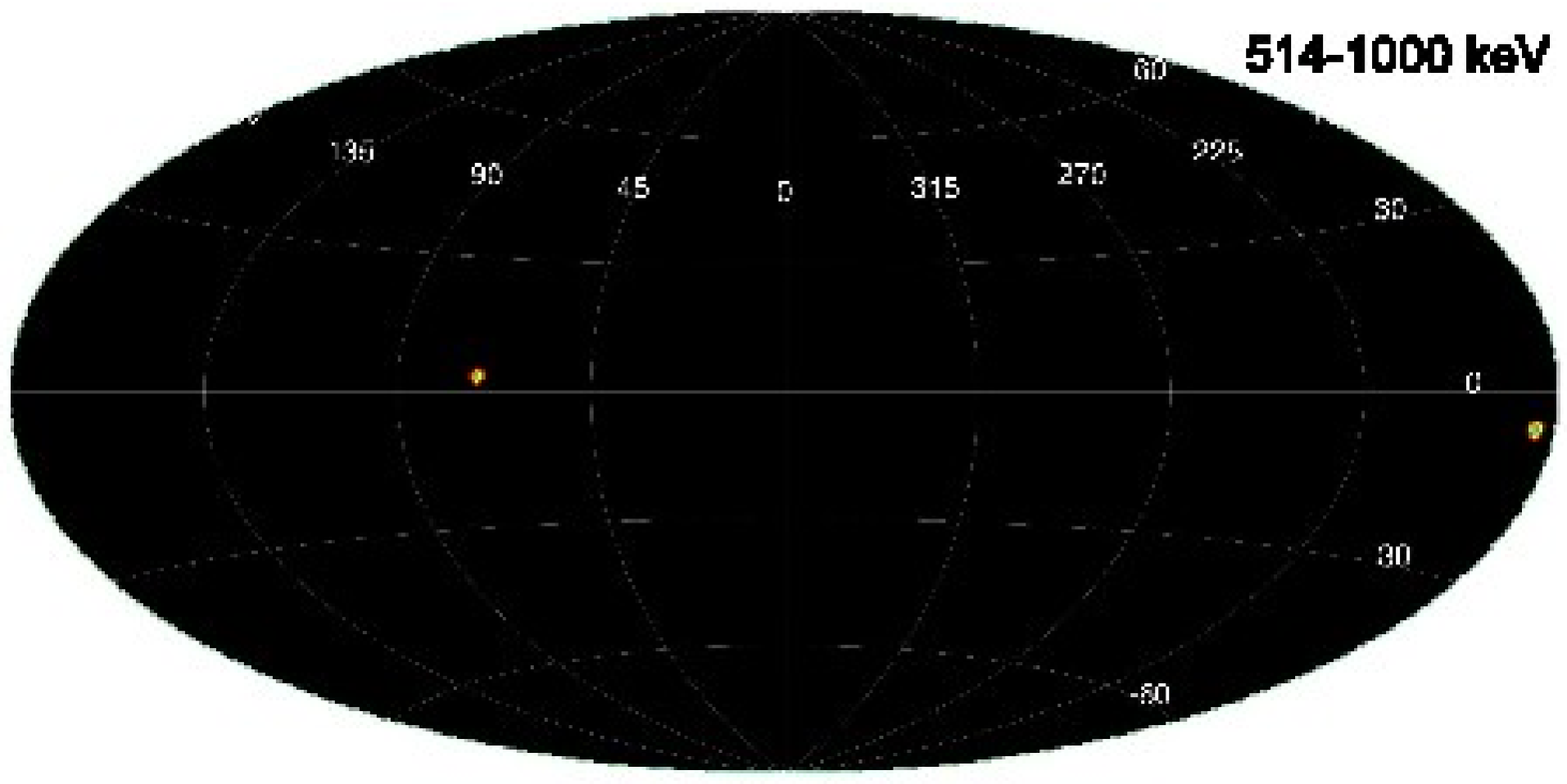}
\includegraphics[width=0.86\linewidth]{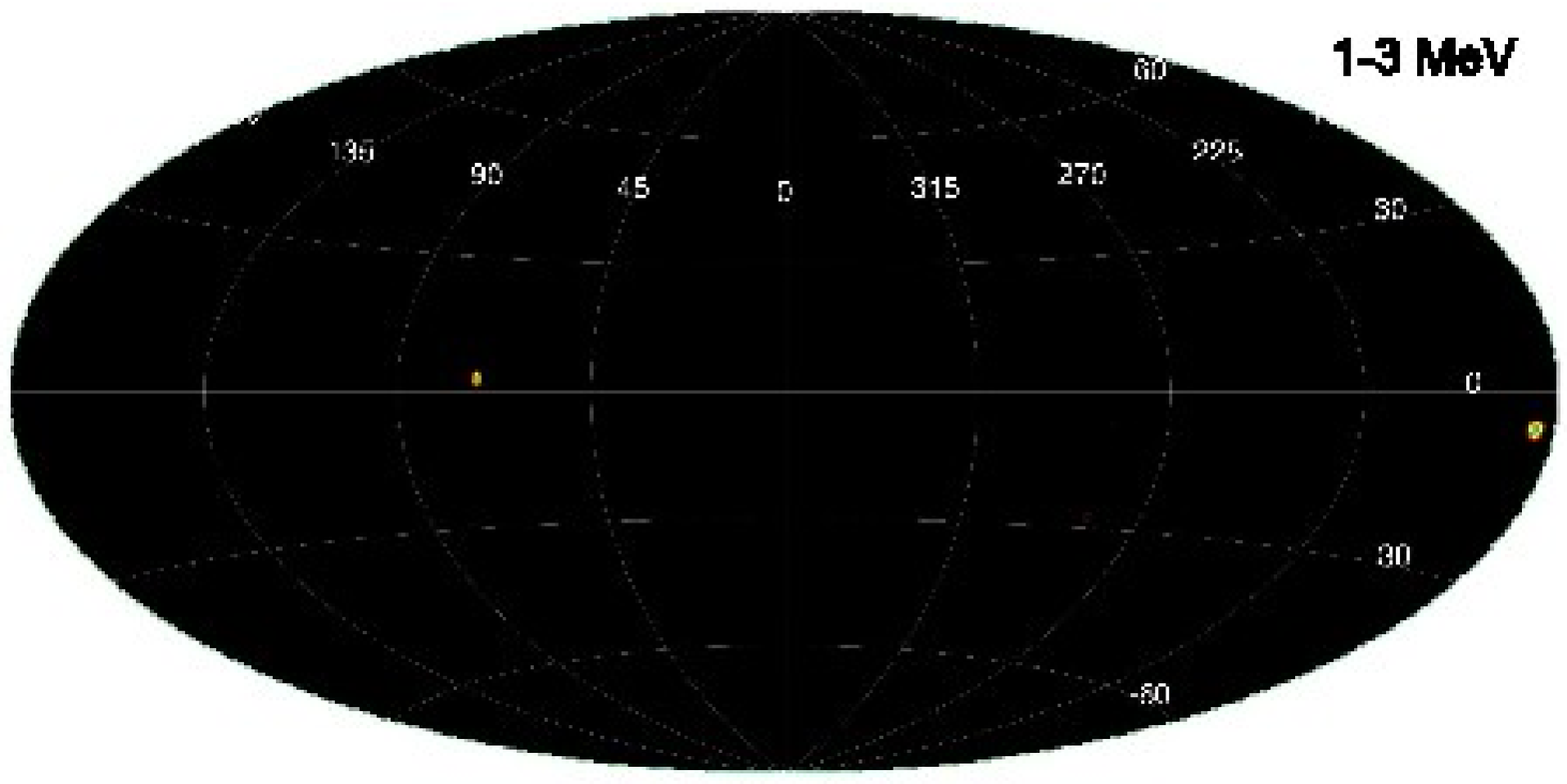}
\caption{
  MREM images of the gamma-ray sky in the continuum energy bands
  514--1000 keV, and 1--3 \MeV.
  \label{fig:cont3}}
\end{figure*}

Our second goal is the imaging of the hard X-ray and soft gamma-ray 
emission in broad continuum energy bands.
This energy range is a transition region where at least three 
emission components are present:
(1) soft ($\la$300~keV) point source emission from Galactic X-ray 
binaries and AGNs,
(2) diffuse bulge dominated emission up to 511~keV attributed to 
ortho-positronium annihilation, and
(3) hard ($\ga$300~keV) extended Galactic plane emission of yet 
unknown nature, possibly linked to the interaction of cosmic-rays with 
the interstellar medium.
The spectral characteristics of these 3 components have been already
extensively studied with SPI \cite{bouchet05, strong05}, but images of 
this transition region, in particular above $\ga$100~keV have so far only 
been published for the ortho-positronium annihilation component
\cite{weidenspointner06}.

In Figs.~\ref{fig:cont1}--\ref{fig:cont3} we present the MREM 
reconstructions for 8 energy bands situated between 20~keV and 3~\MeV. 
The hard X-ray images below 100~keV nicely show the population of 
Galactic point sources, mainly X-ray binaries, of which we detect 
$\sim$60 in the images below 50~keV and $\sim$25 in the 50--100~keV
image.
No diffuse emission is visible in the reconstructions, indicating 
that, if present, such emission should be at a rather low intensity 
level.
In the 100--200~keV image we still count $\sim$18 point sources in 
the image, without a clear hint for diffuse emission.
In the 200--300~keV image, a diffuse and structured emission band along 
the Galactic plane replaces the region where the point sources were 
located at lower energies.
Emission maxima coincide well with the positions of point sources 
found in the 100--200~keV image, and we therefore suggest that at 
least part of the 200--300~keV emission originates from faint and 
unresolved point sources.
We cannot exclude, however, that a substantial fraction of the 
emission in this energy band is indeed of diffuse nature.

The morphology of the emission changes drastically above 300 keV.
The 300--500 keV image is dominated by Galactic bulge emission, 
which is very similar in morphology to the emission that is observed 
in the 511~keV gamma-ray line (see Weidenspointner et al., these 
proceedings).
The only point sources that are still clearly present at these 
energies are the Crab (near the anticentre) and the X-ray binary 
Cyg~X-1.
The bulge emission seems to show some flattening which may be taken as an 
indication of another, yet weaker, underlying emission component that 
follows the Galactic plane.

\begin{figure*}[!ht]
\centering
\includegraphics[width=0.86\linewidth]{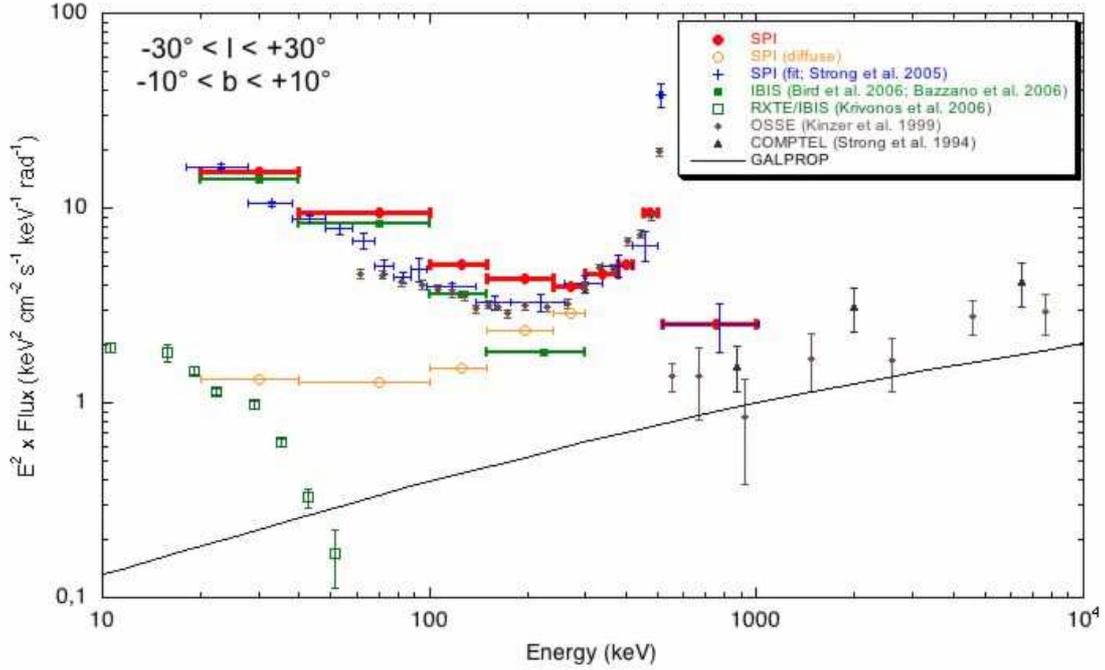}
\caption{
  Comparison of the MREM inner Galactic radian spectrum (filled dots) to 
  SPI model fitting results (crosses, \cite{strong05}) and to spectra 
  obtained by other instruments.
  The filled squares show the summed point source contributions 
  from the second IBIS catalogue \cite{bird06,bazzano06}, normalised to the 
  same Crab spectrum as used for the SPI data.
  The open dots show the residual MREM flux after subtraction of the 
  IBIS point source contribution.
  The open squares show the diffuse hard X-ray emission found by 
  \cite{krivonos06} using IBIS/ISGRI as a light-bucket.
  The triangles show the result obtained from COMPTEL data 
  \cite{strong94}, 
  the diamonds show the OSSE results \cite{kinzer99}.
  The black line is the most recent GALPROP model prediction for 
  galactic diffuse emission (Strong, private communication).
  \label{fig:spectrum}}
\end{figure*}

Above 514 keV (i.e.~at energies situated above the 511 keV 
positron-electron annihilation line), the gamma-ray sky becomes 
rather dark (cf.~Fig.~\ref{fig:cont3}).
Only 2 point sources are visible in the 514--1000~keV and 1--3~\MeV\ 
images, the Crab and Cyg~X-1, while no emission can be seen from the 
Galactic plane.
If we would plot the images to very low intensities we would start to 
see some irregular emission correlated with the Galactic disk, but 
MREM did not manage to extract this component clearly from the data.
In other words, the extended Galactic disk emission that we know from 
spectral analysis to be present above 514~keV in our SPI 
observations \cite{bouchet05, strong05} is too weak to be reliably 
imaged with the present dataset.

However, we can determine the global intensity of the Galaxy at these 
energies by integrating the flux that is present in the MREM images over 
large regions.
We show the result of such an integration in Fig.~\ref{fig:spectrum}.
In order to determine the global flux from the inner Galactic radian, 
we have chosen an integration region of $\pm$30\deg\ in Galactic 
longitudes and $\pm$10\deg\ in Galactic latitudes.
We normalised the spectrum to that of the Crab by assuming for the 
latter a broken power law with a break energy of 117~keV, power law 
indices of -2.04 and -2.37 for below and above the break, respectively, 
and a normalisation of 10.78 at 1 keV.
We derived this prescription by fitting SPI data of the Crab over the 
25--500 keV energy band.

For comparison we show in Fig.~\ref{fig:spectrum} also the spectrum 
that we obtained from SPI data using a model fitting procedure 
\cite{strong05}, normalised to the above mentioned Crab spectrum.
The MREM result matches quite nicely the spectrum obtained by model 
fitting.
The advantage over the model fitting approach is that the MREM
results are unbiased towards assumptions about the spatial distribution 
of the emission, and this difference may eventually explain the 
slightly higher fluxes obtained by MREM in the $\sim100-200$ keV range.
The disadvantage is that MREM does not provide an estimate of the 
statistical uncertainties related to the flux measurements.
Hence, both methods are best used in conjunction, providing thus 
reliable estimates of both the spatial and spectral distribution of 
the emission.

We also show in Fig.~\ref{fig:spectrum} the spectrum that is obtained by 
summing the flux of all point sources listed in the second 
IBIS catalogues \cite{bird06,bazzano06} within the integration region.
We again use the Crab normalisation as for the MREM points to convert the 
catalogue fluxes (quoted in mCrab) into physical units.
For energies below 100~keV, the MREM fluxes match fairly well the point 
source fluxes determined with IBIS/ISGRI, indicating that most of the 
emisison in this band can indeed be attributed to point sources
\cite{lebrun04}.
Subtracting the IBIS/ISGRI point source flux from the SPI MREM 
measurements provides an estimate of the unresolved (or diffuse) emission 
component, which is shown as open dots in Fig.~\ref{fig:spectrum}.
Below $\sim40$ keV, this estimate is consistent with the findings of 
\cite{krivonos06} who report the detection of an unresolved emission 
component using IBIS/ISGRI that they attribute to a large population of 
weak CVs (open squares).
Above $\sim40$ keV, our measurements indicate an additional component 
that is not resolved by IBIS/ISGRI into point sources.
While above $\sim300$ keV ortho-positronium annihilation towards the 
galactic bulge may explain partly this discrepancy, the excess 
emission between $\sim40-300$ keV can only be attributed to an 
additional unresolved or diffuse emission component.

We also show in Fig.~\ref{fig:spectrum} spectral points obtained by 
the COMPTEL \cite{strong94} and OSSE \cite{kinzer99} 
telescopes, but we did not attempt a cross-calibration using the Crab 
for these instruments, which may partially explain some apparent
discrepancies (even between both instruments).
Above 511~keV, the galactic emission is believed to originate mainly 
from cosmic-rays interacting with the interstellar medium and the 
interstellar radiation field, and we show in Fig.~\ref{fig:spectrum} 
the latest prediction of the GALPROP cosmic-ray propagation 
model as solid line (Strong, private communication).
Apparently, GALPROP explains relatively well the spectral shape of 
the unresolved (or diffuse) emission component, both below and above 
the $511$ keV annihilation line, yet the flux level is underestimated by 
a factor $\sim2-3$.
Whether this underestimation is due to a yet unrecognised diffuse 
emission mechanism or due to a yet unresolved weak source population 
remains to be seen.

Fig.~\ref{fig:spectrum} illustrates that MREM images may be exploited 
quantitatively to derive source spectra.
However, such an analysis can not replace a detailed model fitting 
approach which allows to assess the significance of emission 
components and to determine the statistical uncertainties in the 
flux estimates.
MREM spectra should therefore be understood as complementary to the 
model fitting approach in that they do not assume a specific spatial 
distribution for the emission components.
MREM images should hint towards the components that are needed to 
satisfactorily describe the data, model fitting should then be used to 
assess how significant these components indeed are.

\section{Conclusions}

The MREM image reconstruction algorithm presents an interesting 
alternative for the deconvolution of SPI hard X-ray and gamma-ray data.
The main features of this method are (1) the suppression of image 
noise in the reconstructions and (2) the convergence towards a 
well-defined solution.
The application of MREM to simulations nicely illustrates that these 
properties are indeed reached by the present implementation.

We employed the algorithm to present SPI images in the 1809~keV 
gamma-ray line from \al\ and in continuum bands covering energies from
20~keV up to 3~\MeV.
The 1809~keV image confirms the basic emission features that have been 
established by COMPTEL, namely the strong Galactic ridge emission from 
the inner Galaxy, and a prominent emission feature in Cygnus.
The continuum images reveal nicely the transition from a point source 
dominated hard X-ray sky to a diffuse emission dominated soft gamma-ray 
sky.
Imaging the sky above 511~keV is still a challenge, even for MREM, 
due to the low intensity of the Galactic emission and the correspondingly 
small signal that is present in the data.
Nevertheless, MREM allowed to derive spectral points for the Galactic 
emission up to 1~\MeV, which are in fairly good agreement with 
previous measurements by OSSE and COMPTEL.

The accumulation of data beyond the 2 years database used for this 
analysis will enable more detailed imaging of hard X-ray and soft 
gamma-ray emission throughout the spectrum that is accessible to SPI.
We expect to image the diffuse Galactic emission that is underlying 
the strong ortho-positronium continuum in the intermediate
300--511~keV band, and we hope to visualise also the spatial 
distribution of Galactic continuum emission above 511~keV.
Imaging those emissions will be crucial for the determination of 
their origin, which theoretically, is still poorly understood 
\cite{strong05}.

\section*{Acknowledgments}

The SPI project has been completed under the responsibility and leadership 
of CNES.
We are grateful to ASI, CEA, CNES, DLR, ESA, INTA, NASA and OSTC for 
support.


\end{document}